\documentclass[3p]{elsarticle}
\usepackage{natbib}

\makeatletter
\def\ps@pprintTitle{
 \let\@oddhead\@empty
 \let\@evenhead\@empty
 \def\@oddfoot{}
 \let\@evenfoot\@oddfoot}
\makeatother

\usepackage{epsfig}
\usepackage{url}
\usepackage{color}

\usepackage{graphics}
\usepackage{axodraw}
\usepackage{inputenc}
\usepackage{xspace}
\usepackage{tabularx}
\usepackage{def} 

\biboptions{sort&compress}

\begin{document}

\begin{frontmatter}
\begin{flushright}
LU TP 12-40\\ MCnet-13-04
\end{flushright}

\title{\boldmath ColorMath - A package for color summed calculations in
SU(\Nc)}

\author{Malin Sj\"odahl}
\address{  Dept. of Astronomy and Theoretical Physics, Lund University\\
 S\"olvegatan 14A, 223\,62~Lund, Sweden\\
}

\begin{abstract}  
A Mathematica package for color summed calculations 
in QCD (SU(\Nc)) is presented. Color contractions of any color amplitude appearing
in QCD may be performed, and the package uses a syntax
which is very similar to how color structure is written 
on paper. It also supports the definition of color vectors
and bases, and special functions such as scalar 
products are defined for such color tensors.
\end{abstract}

\end{frontmatter}

\sloppy
 
\section{Introduction}
\label{sec:intro}

With the LHC follows an increased demand of exact calculations in QCD 
involving many color charged partons. Due to the non-abelian nature of 
QCD this poses a nontrivial computational problem.

This is the issue the Mathematica\textsuperscript{\textregistered}
package ColorMath\footnote{ColorMath is available at
\url{www.thep.lu.se/~malin/ColorMath.html}.}
is set
out to tackle.
The main feature of ColorMath is thus the ability to automatically perform color summed calculations starting from a QCD color structure which is expressed using a syntax very similar to 
how the color structure would have been
written on paper.

ColorMath allows both partial and full dummy index contractions,
and thus deals with color structures having an arbitrary set of free indices.
In this sense it is thus more general than \cite{Hakkinen:1996bb},
on the other hand it is a pure SU($\Nc$) tool, rather than a high 
energy physics general purpose tool, such as
\cite{Alwall:2011uj,Kuipers:2012rf}.
ColorMath works for an arbitrary 
$\Nc\geq2$, and for arbitrary trace convention of the
squared $\SU(\Nc)$ generators, \TR.
It may be used by just giving the color structure in the appropriate form
and afterwards running \mcom{CSimplify} to contract all repeated indices. 
For example, consider 
$q_1\, \qbar_2 \to q_3\,\qbar_4$ 
via gluon exchange in the s- and t- channels. Writing down the amplitude as

\begin{equation} 
\min{Amp}=\min{S}\, \Colt{g}{q1}{q2} \Colt{g}{q4}{q3}+ 
\min{T}\, \Colt{g}{q1}{q3} \Colt{g}{q4}{q2}
\end{equation}
where $\min{S}$ and $\min{T}$ represent some s- and t-channel kinematics,
and $ \Colt{g}{q1}{q2}$ the SU($\Nc$) generator in the fundamental
representation, otherwise typically denoted $(t^g)^{q1}{}_{q2}$,
we may calculate the squared amplitude (defined as in the scalar product
\eqref{eq:scalar_product}) using
\begin{equation}
	\mdef{CSimplify}[\mdef{Conjugate}[\mdef{Amp}]\,\mdef{ReplaceDummyIndices}[\mdef{Amp}]
	],
	\label{eq:qqExample}
\end{equation}
immediately giving the answer
\begin{equation}
\TR^2\frac{(\Nc^2-1)}{\Nc}[(\Nc S-T)S^*+(\Nc T-S)T^*].
\end{equation}

Each squared amplitude can in principle can be calculated like this.
However, ColorMath also facilitates the usage of color vectors and bases
and has special functions for the calculations of scalar products and
gluon exchanges.

This paper, which is intended to be the main reference, is organized 
as follows: First a general introduction to the basic color
building blocks is given in
\secref{sec:QCD_building_blocks}.
In \secref{sec:basic_computational_strategy} the computational strategy is presented,
whereas basic examples are given in \secref{sec:calculations}, and the usage
of vectors and their functions are presented in 
\secref{sec:tensors}. In \secref{sec:validation} some remarks concerning
validation and scalability are made and in \secref{sec:conclusions} conclusions
are drawn.

\section{QCD building blocks}
\label{sec:QCD_building_blocks}

From a color perspective, the QCD Lagrangian is built out of 

\begin{equation}
\label{eq:t_and_f_def}
  \mbox{quark-gluon vertices,} \qquad 
  \epsfig{file=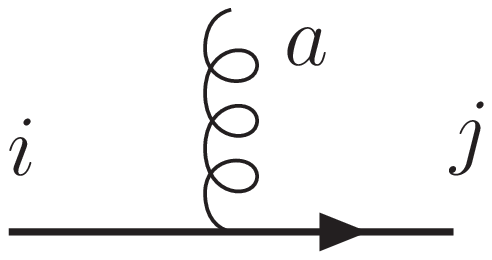,width=2.5cm} = (t^a)^i_{\phantom{i}j},
\end{equation} 
and 
\begin{equation}
  \mbox{triple-gluon vertices,} \qquad 
  \parbox{2.5cm}{\epsfig{file=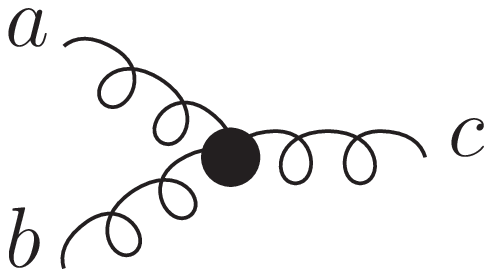,width=2.5cm}} 
  = i f^{abc},
\end{equation} 
where we follow the convention of reading of the fully antisymmetric structure 
constant indices in counterclockwise order. The four gluon vertices can be
rewritten in terms of (one gluon) contracted triple-gluon vertices, and thus need no
special treatment.

Due to confinement, we never observe individual colors and it therefore suffices 
to calculate color summed/averaged quantities for making predictions in
quantum chromodynamics.
We may thus
constrain ourselves to treat QCD amplitudes carrying a set of external indices 
with values that need never be specified, as they are always summed over in the
end.

In principle, the $\SU(\Nc)$ generators, along with a delta function for 
indicating a quark and an anti-quark color singlet,
$\delta^{q1}{}_{q2}$, constitute a minimal set of objects for treating 
the color structure in QCD\footnote{To enhance the similarity with usage inside Mathematica, 
we here use the somewhat unorthodox notation $q1$ etc. 
for single quark and gluon indices.}.
For convenience, and for performance reasons, it is, however, 
useful to define a larger set of objects. 
The complete set of basic building blocks for carrying 
color structure used by the ColorMath package is given in \tabref{tab:basic_objects}.

Apart from a delta function in quarks indices, $\Coldelta{q1}{q2}$,
a delta function in gluon indices, denoted by $\ColDelta{g1}{g2}$, is also defined. 
Note the Mathematica \mcom{List} brackets in {\small $\{{\blue g1},{\blue
g2}\}$}. The gluon delta function can alternatively be expressed using

\begin{equation}
\parbox{2.5cm}{\epsfig{file=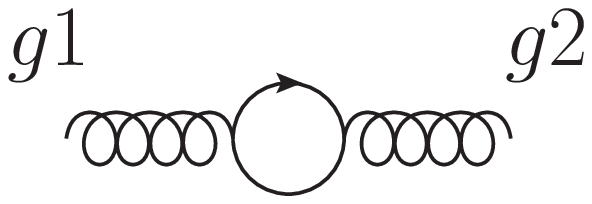,width=2.5cm}}
=\TR \parbox{2cm}{\epsfig{file=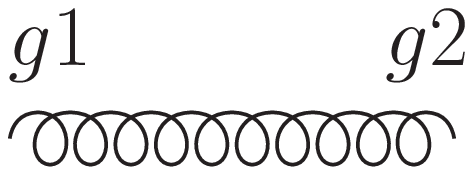,width=2cm}}
\Leftrightarrow\;
\tr[t^{g1} t^{g2}] 
=\TR \,\delta^{g1g2} 
\;\Leftrightarrow\;
\colo^{ \{{\blue g1},{\blue g2}\}}
=\mdef{TR}\,\ColDelta{g1}{g2}
\label{eq:tr}
\end{equation}
where $\TR$ ($\mdef{TR}$) comes from the normalization of the $\SU(\Nc)$
generators, and is typically taken to be $1/2$ (the Gell-Mann normalization) or $1$. 
A rescaling of the normalization of the $\SU(\Nc)$ generators can always be 
absorbed into a normalization of the strong coupling constant.
To allow for arbitrary
normalization $\TR$ is kept as a free parameter, denoted by $\mdef{TR}$, and may
be defined by the user. 
Also the number of colors, denoted $\mdef{Nc}$ in ColorMath, may be set by the
user; by default both $\mdef{Nc}$ and $\mdef{TR}$ are kept as free parameters.
Note the ColorMath notation for a trace over
two gluons $\colo^{ \{{\blue g1},{\blue g2}\}}$ in \eqref{eq:tr}. 
Similarly, a general trace over $k$ gluons {\small $g1,g2,..gk$}
is denoted $\COLo{g1}{g2}{\ldots}{gk}$, and may be thought of as a closed 
quark-line with $k$ gluons attached.

\begin{table}[t]
\caption{\label{tab:basic_objects} Along with the number of colors,
$\mdef{Nc}$, and the trace of an {SU($\Nc$)} generator squared,
$\mdef{TR}$, the basic building blocks are as below.
Note that $\COLO{g1}{g2}{\ldots}{gk}$ represents a trace over gluons with
indices $g1...gk$, $\tr[t^{g1}t^{g2}...t^{gk}]$, and that
$\COLT{g1}{g2}{\ldots}{gk}{q1}{q2}$ represent the $q1,q2$-component in
a trace over generators that has been cut open,
$(t^{g1}t^{g2}...t^{gk})^{q1}{}_{q2}$.
For convenience, also the totally symmetric ``structure constants''
$\Cold{g1}{g2}{g3}$ are defined.
Note the Mathematica \mcom{FullForm}. ColorMath is built on 
pattern matching, and it is therefore essential that the expressions
have the correct \mcom{FullForm}. In particular, \mcom{Power} may
not be used instead of \mcom{Superscript}. To get the right \mcom{FullForm}
it is recommended to use the function form, which is just a function returning
the corresponding ColorMath object.} \vspace*{0.2 cm}
\begin{tabular}{|p{3.3 cm} |p{1.9cm} | l | l |}
\hline 
Pictorial representation &
ColorMath &
Function form & 
Mathematica \mcom{FullForm} \\ [0.5ex] 
\hline 
\epsfig{file=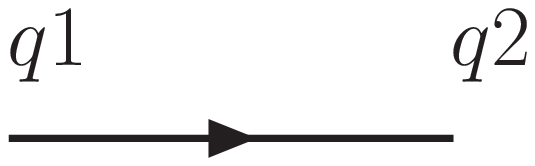,width=2.5cm} & 
$\Coldelta{q1}{q2}$   &  
\coldelta[\min{q1},\,\min{q2}] &
\mcom{Subscript}[\mcom{Superscript}[$\setminus$[\mcom{Delta}],\,q1],\,q2]
\\  
\epsfig{file=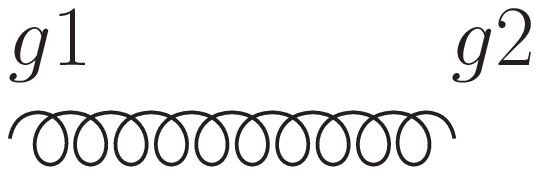,width=2.5cm}  & 
$\ColDelta{g1}{g2}$    &  
\colDelta[\min{g1},\,\min{g2}] &
\mcom{Superscript}[$\setminus$[\mcom{CapitalDelta}],\,\mcom{List}[g1,\,g2]]
\\
$\frac{1}{i}
\parbox{2.5cm}{\epsfig{file=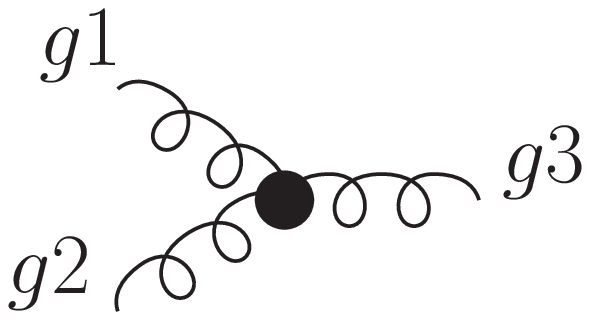,width=2.5cm}}$ &
$\Colf{g1}{g2}{g3}$   & 
\colf[\min{g1},\,\min{g2},\,\min{g3}]  &
\mcom{Superscript}[\colf,\,\mcom{List}[g1,\,g2,\,g3]]
\\
\epsfig{file=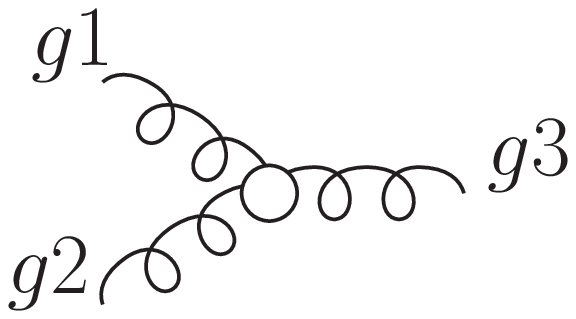,width=2.5cm}&
$\Cold{g1}{g2}{g3}$   &
\cold[\min{g1},\,\min{g2},\,\min{g3}]  &
\mcom{Superscript}[\cold\,,\mcom{List}[g1,\,g2,\,g3]]
\\
\epsfig{file=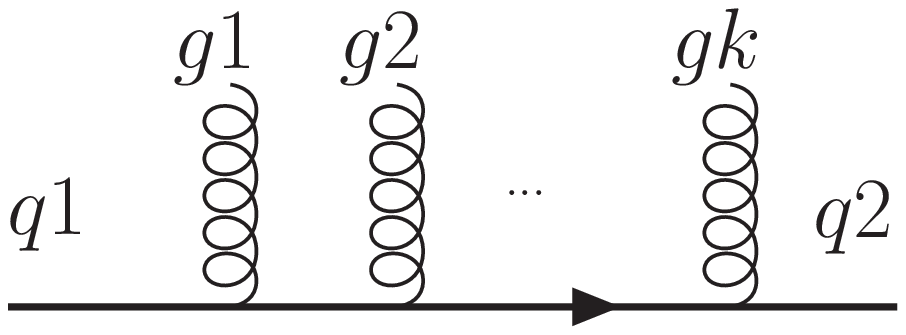,width=3.5cm}&
$\COLt{g1}{\ldots}{gk}{q1}{q2}$  &
\colt[\{\min{g1},\ldots,\,\min{gk}\},\,\min{q1},\,\min{q2}]   &
\begin{tabular} {l}
\mcom{Subscript}$[$\mcom{Superscript}$[$\mcom{Superscript}$[$\colt, \\
\mcom{List}[g1,\ldots,\,gk]],\,q1],\,q2]
\end{tabular}
\\ 
\epsfig{file=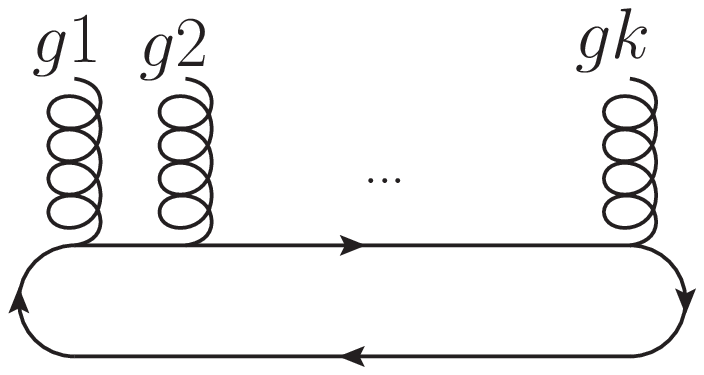,width=3.5cm}&
$\COlo{g1}{\ldots}{gk}$ &
\colo[\{\min{g1},\ldots,\,\min{gk}\}]  &
\mcom{Superscript}[\colo,\,\mcom{List}[g1,\,\ldots ,\,gk]]
\\ \hline
\end{tabular} 
\end{table}

The totally antisymmetric structure constants, 
which -- along with an extra $i$ -- define the triple gluon vertices, are
denoted $\Colf{g1}{g2}{g3}$. Similarly, the totally symmetric ``structure
constants'' are defined as $\Cold{g1}{g2}{g3}$. Recall that, starting from the
commutation (anticommutation) relations

\begin{eqnarray}
[t^{g1},t^{g2}]= i f^{g1\,g2\,g3} \,t^{g3} \;,\; \quad 
\{t^{g1},t^{g2}\}= d^{g1\,g2\,g3} \,t^{g3},
\end{eqnarray}
the structure constants can be rewritten in terms of
traces over $\SU(\Nc)$ generators,
\begin{eqnarray}
  \label{eq:fd}
  i f^{g1\,g2\,g3}/d^{g1\,g2\,g3} &=& \frac{1}{\TR} \big[ \tr(t^{g1} t^{g2}
  t^{g3}) \mp \tr(t^{g2} t^{g1} t^{g3}) \big]\\
  &=&\frac{1}{\TR} 
  \big[
    (t^{g1})^{q1}{}_{q2}  (t^{g2})^{q2}{}_{q3}  (t^{g3})^{q3}{}_{q1} 
    \mp
     (t^{g2})^{q1}{}_{q2}  (t^{g1})^{q2}{}_{q3}  (t^{g3})^{q3}{}_{q1}
     \big].\nonumber
\end{eqnarray}
In ColorMath notation $\colt$ is used to denote an open 
quark-line, and the above expression is written similarly, 
\begin{eqnarray}
\label{eq:fdMat}
  \colI \, \Colf{g1}{g2}{g3}/ \Cold{g1}{g2}{g3} 
  &=&  \frac{1}{\mdef{TR}} \big(\Colo{g1}{g2}{g3} \mp \Colo{g2}{g1}{g3}  
  \big)\\
  &=&\frac{1}{\mdef{TR}} 
  \big(
    \Colt{g1}{q1}{q2}\; \Colt{g2}{q2}{q3} \;\Colt{g3}{q3}{q1}
    \mp 
    \Colt{g2}{q1}{q2} \;\Colt{g1}{q2}{q3}\; \Colt{g3}{q3}{q1}
    \big).\nonumber
\end{eqnarray}
 Pictorially this represents

\begin{eqnarray}
  \label{eq:fdPic}
  \parbox{1.5cm}{\epsfig{file=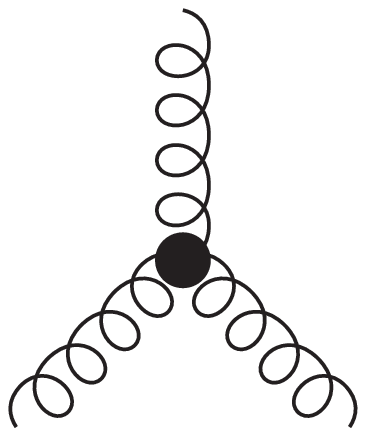,width=1.5cm}} &=&
  \frac{1}{\TR} \left[ 
    \parbox{1.5cm}{\epsfig{file=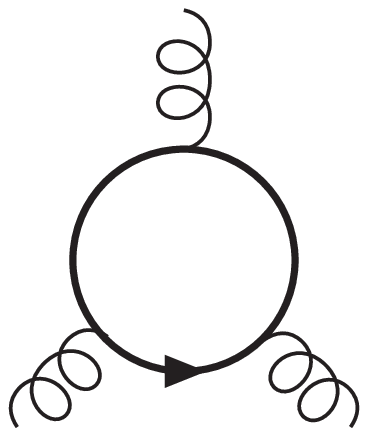,width=1.5cm}}
    - \parbox{1.5cm}{\epsfig{file=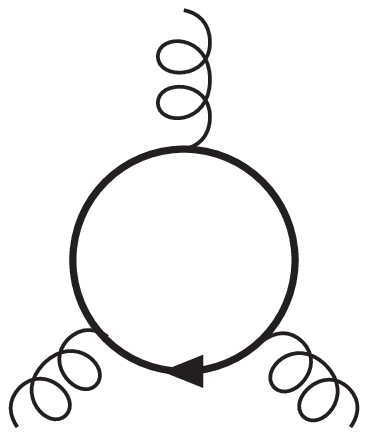,width=1.5cm}} \right]
  \;,\; \quad 
  \parbox{1.5cm}{\epsfig{file=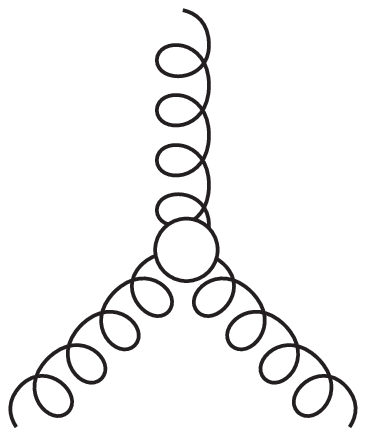,width=1.5cm}} =
  \frac{1}{\TR} \left[ 
    \parbox{1.5cm}{\epsfig{file=Figures/f_via_3t_2,width=1.5cm}}
    + \parbox{1.5cm}{\epsfig{file=Figures/f_via_3t_3,width=1.5cm}}
    \right],
\end{eqnarray}
where $i$ is included in the triple gluon vertex.
The rationale for putting the gluon index in the SU($\Nc$) generators 
$ \Colt{g2}{q1}{q2}$ inside a 
Mathematica \mcom{List} in \eqref{eq:fdMat} is to allow for the natural
extension of having many gluons attached to an open quark-line, thus 

\begin{eqnarray}
\COLT{g1}{g2}{...}{gk}{q1}{q2}=
\Colt{g1}{q1}{d2} \Colt{g2}{d2}{d3}....\Colt{gk}{dk}{q2}.
\end{eqnarray}
The left hand side has several advantages compared to the right hand side.
An open quark-line with an arbitrary number of gluon indices 
can be written in a compact form with no dummy indices. 
This is not only more human readable, but also superior from 
a computational point of view, as it avoids the contraction of unnecessary
dummy indices.
A third advantage with the above notation is its direct correspondence to the 
trace type bases
\cite{Paton:1969je,Dittner:1972hm,Cvi76,Cvitanovic:1980bu,Mangano:1987xk,Mangano:1988kk,
Nagy:2007ty,Sjodahl:2009wx,Platzer:2012np}.
A basis (or spanning set) for the color space for a fixed set of external quarks, 
anti-quarks and gluons can always be 
taken to be a sum of products of open and closed quark-lines.

In this context we also remark that the color tensors defined in 
\tabref{tab:basic_objects}
are color scalars, i.e., they are invariant under SU(\Nc) transformations.
This imposes no restriction for our purposes as, for any QCD amplitude, the
overall color structure, including {\it both} incoming and outgoing particles,
always is a color singlet.
As the basic
building blocks are invariant, each tensor built out of these objects, 
i.e., each tensor needed for color summed calculations in 
perturbative QCD is a actually a color scalar.

The scalar product on this vector space is given by summing over all external color
indices, i.e.
\begin{equation}
  \left\langle \Col_1 | \Col_2 \right\rangle
  =\sum_{a_1,\,a_2,\,...}\Col_1^{* a_1\,a_2...} \, \Col_2^{a_1\,a_2...} 
\label{eq:scalar_product}
\end{equation}
with $a_i=1,...,\Nc$ if parton $i$ is a quark or anti-quark and
$a_i=1,...,\Nc^2-1$ if parton $i$ is a gluon. As long as the 
color structures in \tabref{tab:basic_objects} are multiplied 
by real coefficients the scalar product is actually real,
which is easy to prove using the computational rules in the next section.

\section{Basic computational strategy}
\label{sec:basic_computational_strategy}

Having defined all the color carrying objects, we turn to describing the basic 
strategy for carrying out calculations.

Again we note that we need not treat the four gluon vertex as this 
can be rewritten in terms of three gluon vertices.
To treat an arbitrary color structure in QCD we may always compute any 
squared amplitude in the following way:

\begin{enumerate}
\item[(i)]{
Rewrite the triple gluon vertices using \eqref{eq:fd}. 
This results in a color structure 
which is a sum of products of open and closed quark-lines, 
connected to each other via repeated gluon indices.}
\item[(ii)]{Contract all internal gluon indices using the Fierz or 
completeness relation
\begin{eqnarray}
  \parbox{2cm}{\epsfig{file=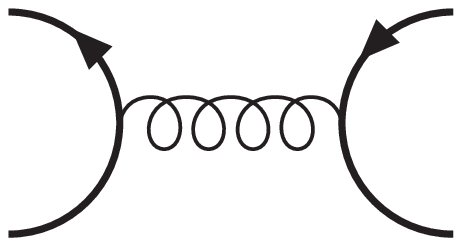,width=2cm}}
  &=&\TR \left[ 
    \parbox{1.5cm}{\epsfig{file=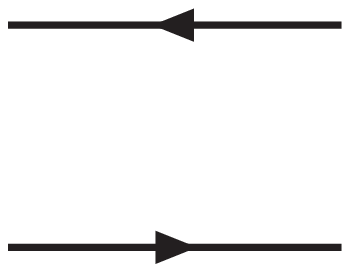,width=1.5cm}}
    - \frac{1}{\Nc} 
      \parbox{1.5cm}{\epsfig{file=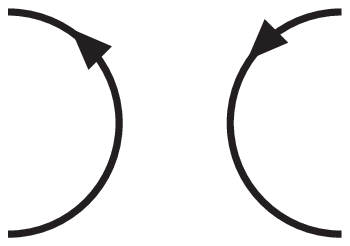,width=1.5cm}} \right].
\label{eq:Fierz}
\end{eqnarray}
}
\item[(iii)]{If present, remove Kronecker deltas (for quarks and gluons), 
and internal quark indices using 
\begin{eqnarray}
 \parbox{7.8cm}{\epsfig{file=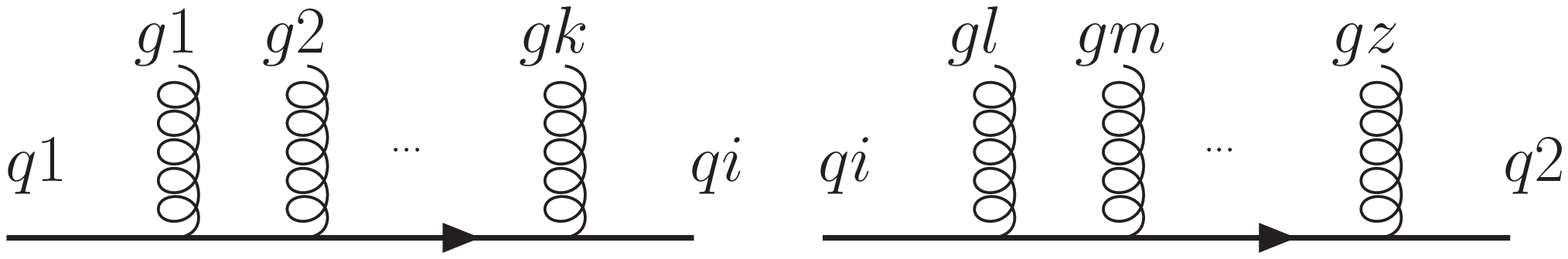,width=7.8cm}}
 &=&
 \parbox{5.1cm}{\epsfig{file=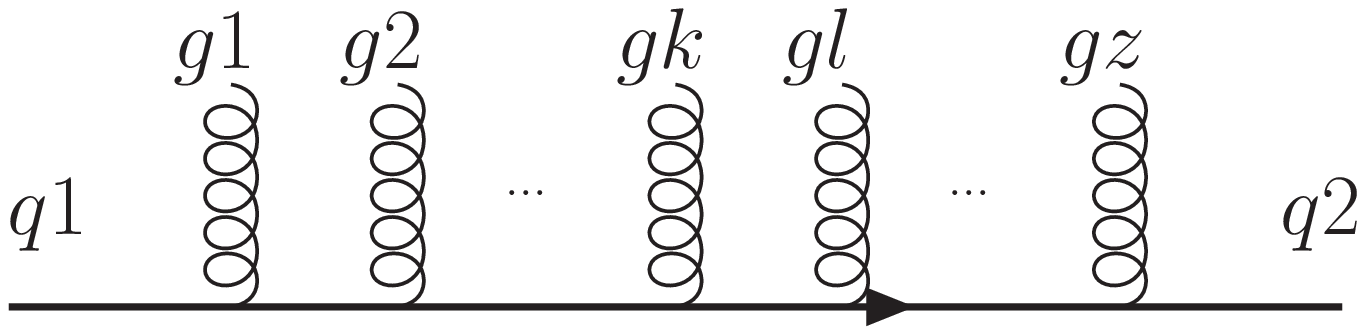,width=5.1cm}}.
\end{eqnarray}
}
\item[(iv)]{ Use the tracelessness of the SU(\Nc) generators, and contract quark
and gluon delta functions with repeated indices

\begin{eqnarray}
\label{eq:delta_cont}
  \parbox{2.2cm}{\epsfig{file=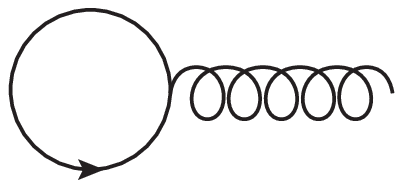, width=2.2cm}}&=&0 \nonumber \\
  \parbox{1.4cm}{\epsfig{file=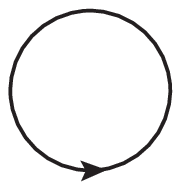, width=1.4cm}}&=&\Nc
  \nonumber\\
  \parbox{1.7cm}{\epsfig{file=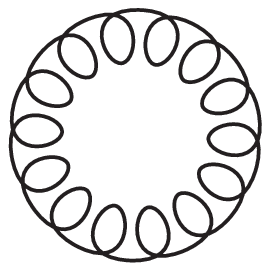,width=1.7cm}}&=& \Nc^2-1.
\end{eqnarray} 
}
\end{enumerate}
After this, all dummy indices have been contracted. A squared amplitude is thus
represented by a rational function in $\TR$ and $\Nc$, whereas a
general amplitude is represented as a sum of products of closed and open
quark-lines, i.e., as sums of products of $\COLT{g1}{g2}{...}{gk}{q1}{q2}$, including  
$\Colt{}{q1}{q2}=\Coldelta{q1}{q2}$,  
and $\COLo{g1}{g2}{...}{gk}$, which for two gluons may be rewritten 
as $\colo^{\{ {\blue g1},{ \blue g2} \} }=\mdef{TR}\,\ColDelta{g1}{g2}$.

By applying these rules, {\it any} amplitude square
and any interference term appearing in QCD can be calculated  \cite{Cvi76,
Sjodahl:2009wx}.
To successfully square an arbitrary QCD amplitude using the above set of rules
we see that steps (i-ii) have to be performed while keeping the relative order,
i.e., by first applying rule (i) and then rule (ii).
We also note that these rules increase the number of terms, whereas the rules
(iii-iv) decrease the number of terms or keep it fixed.
In order not to unnecessarily inflate an expression it may therefore be useful to 
apply the latter rules at any time during the computation. ColorMath utilizes
this and tries to contract indices using (iii-iv) at any time, while rules
(i-ii) are used only when needed, i.e., when the non-expanding rules fail.

On top of the rules (iii-iv) there are other cases in which the contraction of a
gluon results in at most one term. This will happen if 

\begin{enumerate}
\item[(v)]{Two neighboring gluons attached to the same (closed or open)
quark-line are contracted. In this case the result is simply $\cf=\TR(\Nc^2-1)/\Nc$ times the color
structure where the two involved gluons have been contracted,
\begin{eqnarray}
 \parbox{5.8cm}{\epsfig{file=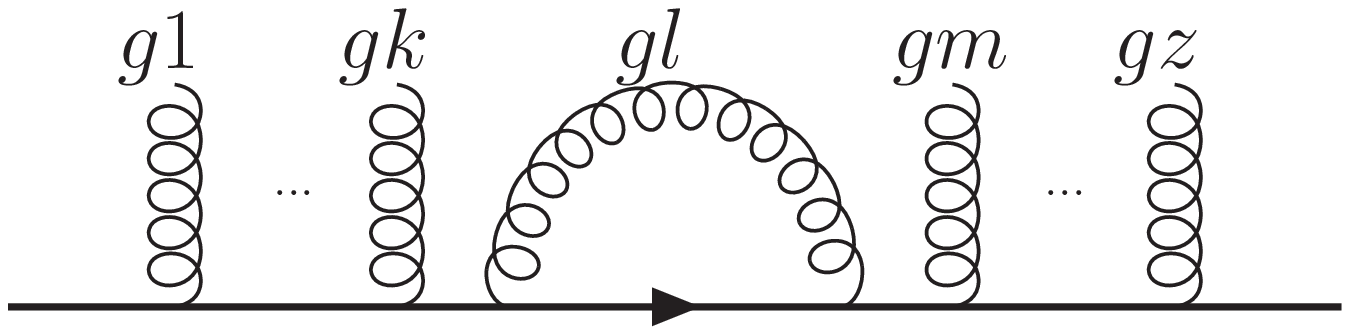,width=5.8cm}}
 = \TR \frac{\Nc^2-1}{\Nc}\parbox{3.5cm}{\epsfig{file=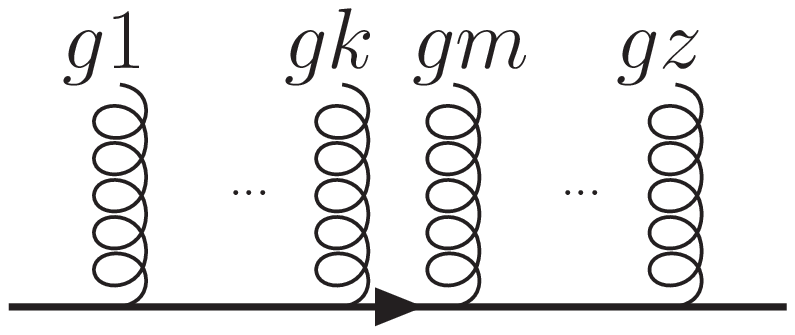,width=3.5cm}}.
\label{eq:neighbors}
\end{eqnarray}
}

\item[(vi)]{Similarly, it is easy to show, using the Fierz identity,
\eqref{eq:Fierz}, that the contraction of two next to neighboring gluons 
results in only one term

\begin{eqnarray}
 \parbox{5.5cm}{\epsfig{file=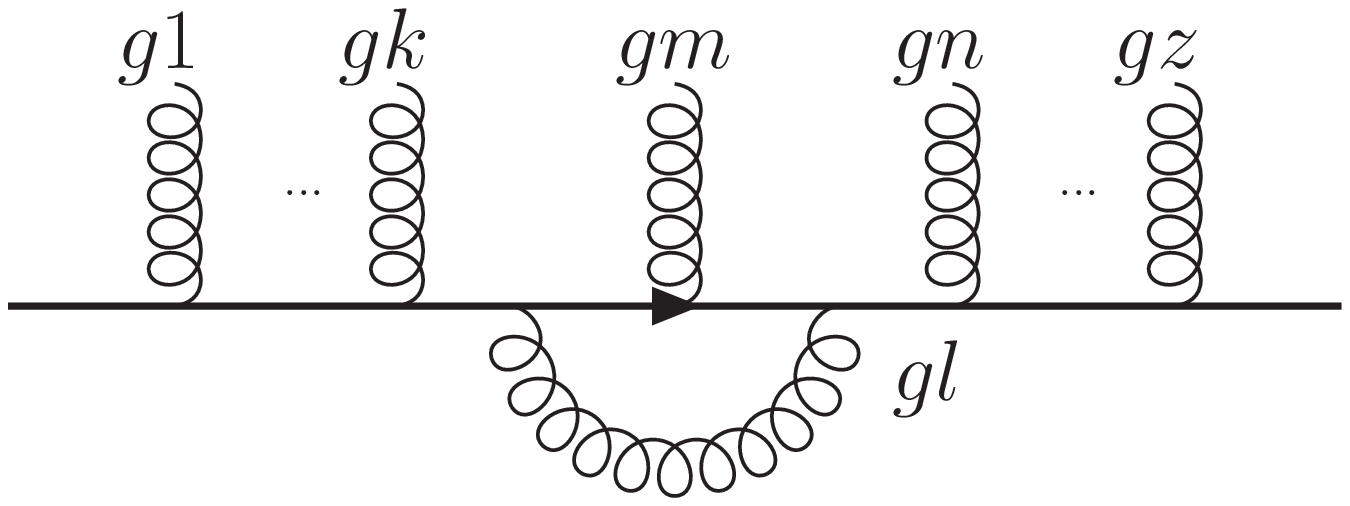,width=5.5cm}}
 = -\TR \frac{1}{\Nc}\parbox{5.5cm}{\epsfig{file=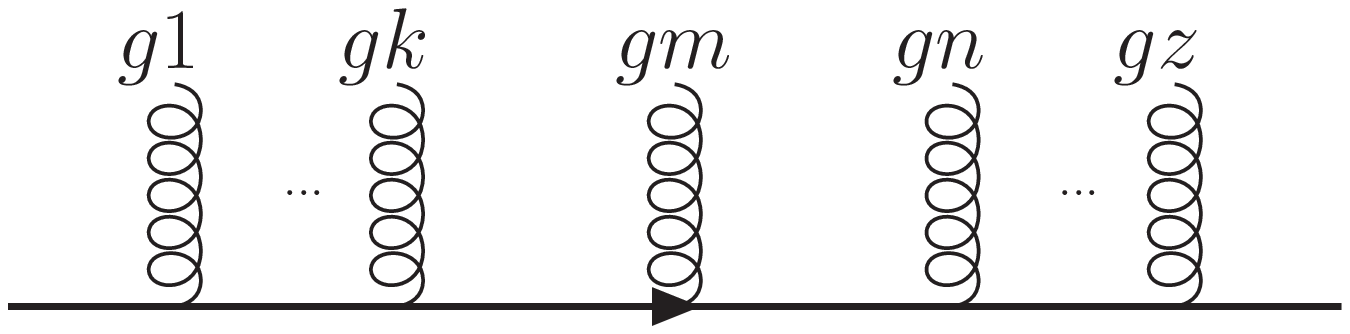,width=5.5cm}}.
\label{eq:next_neighbors}
\end{eqnarray}
}

\end{enumerate}

While the above rules are sufficient for squaring color
amplitudes, it is sometimes advantageous to contract indices 
directly using the structure constants $f^{g1\,g2\,g3}$ and $d^{g1\,g2\,g3}$.
Therefore rule (i) is {\it not} always automatically used for simplifying 
expressions. Instead, in addition to the above set of rules, 
a limited set of rules for contraction of repeated gluon indices 
occurring in the standard and the symmetric structure constants are implemented.
More specifically, all rules for contracting two gluons in products of 
two symmetric or antisymmetric structure constants, for example,

\begin{eqnarray}
    f^{g1\,i1\,i2} f^{g2\,i1\,i2} = 2 \Nc \, \TR\, \delta^{g1\,g2}
  \label{eq:fd1}
\end{eqnarray}
and all rules for contracting three gluons in products
of three symmetric or antisymmetric structure constants, such as

\begin{eqnarray}
 d^{g1\,i1\,i2} d^{g2\,i1\,i3} f^{g3\,i2\,i3}=
 \TR \frac{\Nc^2-4}{\Nc} f^{g1\,g2\,g3}
  \label{eq:fd2}
\end{eqnarray}
are implemented.
For more intricate gluon contractions involving structure 
constants ColorMath can contract indices by first applying rule (i).

\section{Basic calculations}
\label{sec:calculations}

In this section, basic functionality of ColorMath is explored.
As always with Mathematica, the package has to be loaded before it can be used. 
With 

\begin{equation}
\mdef{Get}[\mbox{"{\gray/full/path/to/ColorMath1.0.m}"}]
\end{equation}
where the version number is adjusted to the version in question, this can be
done.
To start using the package, it is, however, recommended to start from (a
version of) the tutorial ``ColorMathTutorial.nb'' and modify its content
according to the desired usage.

The color contractions corresponding to the basic manipulations from
\secref{sec:basic_computational_strategy} are carried out using Mathematica \mcom{Rules}, 
i.e., a set of replacement rules based on pattern matching. As always, the rules
may be applied using ``$\mdef{Expr/.TheRules}$'', and may be applied repeatedly using 
``$\mdef{Expr//.TheRules}$''. 
 
The rules described in (iii-iv) are, along with \eqref{eq:tr} 
contained in a set of
rules called \mcom{SimpleRules}, as they keep the expression at least as simple (in a term counting sense) as it initially was.
Applying these to an expression thus tend to simplify it, for example
\begin{eqnarray}
\Colt{g1}{q1}{q2}\,\Colt{g2}{q2}{q1}\,\ColDelta{g1}{g2}/.\mdef{SimpleRules}
\end{eqnarray}
results in $ \mdef{TR} (\ColDelta{g1}{g2})^2$. 

To fully utilize the rules we (may) need to apply them repeatedly,
\begin{eqnarray}
\Colt{g1}{q1}{q2}\,\Colt{g2}{q2}{q1}\,\ColDelta{g1}{g2}//.\mdef{SimpleRules}
\end{eqnarray}
giving the fully contracted expression $(-1+\mdef{Nc}^2)\mdef{TR}$.

The rules defined in (v-vi), acting on $\colt$ and $\colo$, are similarly
contained in
\mcom{OTSimpleRules}, and the union of \mcom{SimpleRules} and
\mcom{OTSimpleRules} and a few rules for rewriting closed quark-lines
with zero to two gluons, are contained in \mcom{AllSimpleRules}.

The special rules for gluon index contraction in structure constants,
exemplified in equation (\ref{eq:fd1}-\ref{eq:fd2}) are defined in \mcom{FDRules}.
The complete set of rules are stated in \tabref{tab:rules}.

Rather than thinking about how individual rules have to be applied, it is
convenient to have a standard procedure for contracting color indices.
This is embodied in the function \mcom{CSimplify}, which does what its name
suggests; simplifies the color structure as far as possible. For color
structures which do not contain structure constants this always implies
contracting all repeated indices.

For color structure involving the structure constants \mcom{CSimplify} 
first attempts simplification using the \mcom{FDRules}. 
If, after this, the expression still contains structure constants, 
the structure constants are by default
rewritten in terms of traces using \eqref{eq:fdMat}, and the indices are fully
contracted, resulting in a sum of products of open and closed quark-lines. 
Sometimes it may, however, be desirable {\it not} to rewrite the structure
constants, as expressions may be more compact if they are kept.
This can be achieved by using the option \mcom{RemoveFD}$\to$ \mcom{False}.

The most useful set of functions are given in \tabref{tab:functions}. 
Apart from \mcom{CSimplify} we especially note the function
\mcom{ReplaceDummyIndices} for replacing all repeated indices 
in a color structure with a new unique set of color indices.

For the purpose of calculating amplitudes square, we need, as in 
\eqref{eq:scalar_product}, the complex conjugated version of the color
structures.
We note that 

\begin{eqnarray}
& &\mdef{Conjugate}[\Coldelta{q1}{q2}]=\Coldelta{q2}{q1}\nonumber \\
&
&\mdef{Conjugate}[\COLt{g1}{...}{gk}{q1}{q2}]=\COLt{gk}{...}{g1}{q2}{q1}\nonumber
\\
& &\mdef{Conjugate}[\Colo{g1}{...}{gk}]=\Colo{gk}{...}{g1} 
\end{eqnarray} 
whereas $\ColDelta{g1}{g2}$, $\Colf{g1}{g2}{g3}$ and
$\Cold{g1}{g2}{g3}$ are real.
This is implemented in ColorMath via redefinition of the Mathematica's built
in function \mcom{Conjugate}. With this in mind, we are ready to 
perform calculations of the type in \eqref{eq:qqExample}.

Finally we remark that to Mathematica 
$\ColDelta{g1}{g2} \ne \ColDelta{g2}{g1}$, and similarly,
$\COLO{g2}{g3}{g4}{g1} \ne \COLO{g1}{g2}{g3}{g4}$ etc. For this reason a
function \mcom{SortIndices}, which writes indices in Mathematica default order is defined, s.t. for
example 

\begin{eqnarray}
  \mdef{SortIndices}[\COLO{g2}{g3}{g4}{g1}]=\COLO{g1}{g2}{g3}{g4}.
\end{eqnarray} 
This function is used by \mcom{CSimplify}, and by some of the rules in
\tabref{tab:rules}.

Sometimes more detailed control over the calculation may be desired.
For this purpose, and for internal usage, functions manipulating indices 
and probing the color structure are given in \tabref{tab:index_functions},
in \ref{sec:index_functions}.

While each squared amplitude in principle can be calculated as outlined above,
ColorMath also offers more efficient tools for dealing with vectors in color
space.

\section{Defining and using vectors}
\label{sec:tensors}

For the purpose of studying color space it is often convenient to define 
a basis for the color space, and sometimes also projection operators.
Both of these are examples of color (singlet) tensors, and ColorMath has
a set of tools for working directly with such tensors.

As an example, let us consider the color structure for 
$q_1 \qbar_2 \to q_3 \qbar_4 g_5$\footnote{In this user guide indices
representing incoming quarks and outgoing anti-quarks are placed upstairs,
whereas outgoing quarks and incoming anti-quarks are placed downstairs. Note, however, that we 
could as well have used the opposite convention.}. A basis for the color space
may be written as \cite{Sjodahl:2008fz}
 
\begin{eqnarray}  
  & &\min{Vector181}
  	_{\{ \mpat{q1\_} ,\mpat{q2\_},\mpat{q3\_},\mpat{q4\_},\mpat{g5\_} \}}:=
  \coldelta^{\mpat{q1}}{}_{\mpat{q2}}\,\colt^{\{\mpat{g5}\}\mpat{q4}}{}_{\mpat{q3}}
  \nonumber
  \\
  & &\min{Vector818}_{\{ \mpat{q1\_} ,\mpat{q2\_},\mpat{q3\_},\mpat{q4\_},\mpat{g5\_} \}}
  	:= \colt^{\{\mpat{g5}\}\mpat{q1}}{}_{\mpat{q2}}
  \,\coldelta^{\mpat{q4}}{}_{\mpat{q3}} \nonumber \\
  & &\min{Vector888s}_{\{ \mpat{q1\_} ,\mpat{q2\_},\mpat{q3\_},\mpat{q4\_},\mpat{g5\_} \}}:= 
  \mdef{Module}[\{\mmod{i1},\mmod{i2}
  \},\colt^{\{\mmod{i1}\}\mpat{q1}}{}_{\mpat{q2 }}
  \,\colt^{\{\mmod{i2}\}\mpat{q4}}{}_{\mpat{q3}}\,
  \cold^{\{\mmod{i1},\mmod{i2},\mpat{g5}\}}]
  \nonumber
  \\
  &  &\min{Vector888a}_{\{ \mpat{q1\_} ,\mpat{q2\_},\mpat{q3\_},\mpat{q4\_},\mpat{g5\_} \}}:=
    \mdef{Module}[\{\mmod{i1},\mmod{i2}
  \},\colt^{\{\mmod{i1}\}\mpat{q1}}{}_{\mpat{q2 }}
  \,\colt^{\{\mmod{i2}\}\mpat{q4}}{}_{\mpat{q3}}\,
  \colI \, \colf^{\{\mmod{i1},\mmod{i2},\mpat{g5}\}}]
  \label{eq:2qqbarg} 
\end{eqnarray}
where the basis vectors are labeled using first the overall multiplet of $q_1 \qbar_2$,
then the overall multiplet of $q_3 \qbar_4$, and finally the overall
multiplet of $q_3 \qbar_4 g_5$, which -- due to color conservation --
must equal the multiplet of $q_1 \qbar_2$, 
(implying that the notation is somewhat redundant).

From a Mathematica perspective we note a few things. First, on the left hand 
side, we see that the indices inside the \mcom{List} in the \mcom{Subscript} 
are followed by underscore to indicate pattern matching. 
This is standard in Mathematica and makes it possible to use any symbol to
denote the indices in later calculations.
Then we note that the last two tensors in \eqref{eq:2qqbarg} are defined using \mcom{Module}. This is to ensure that each
time the tensor is used, it comes with a fresh set of dummy indices.
This is also the reason why set delayed ``:='' is used.

This basis is orthogonal since at least one set of partons transform under
different representations in the various tensors. It is, however, not
normalized. Finding the normalization is easy using ColorMath.
To calculate for example $\min{Vector181}$ squared we could enter
\begin{eqnarray}
 \mdef{CSimplify}[
 \mdef{Conjugate}[ \min{Vector181}_{\{
  \mbox{\small \blue q1},\mbox{\small \blue  q2},\mbox{\small \blue 
  q3},\mbox{\small \blue q4},\mbox{\small \blue g5} \}}]\,
 \min{Vector181}_{\{ \mbox{\small \blue q1},\mbox{\small \blue
 q2},\mbox{\small \blue q3},\mbox{\small \blue q4},\mbox{\small \blue g5} \}} ],
\end{eqnarray}
but it is yet much easier to use the tensor functions for calculating scalar
products.
Instead we could simply write 

\begin{eqnarray}
 \mdef{CDot}[\min{Vector181}, \min{Vector181}]
\end{eqnarray}
resulting in $ \mbox{Nc} \left(-1+\mbox{Nc}^2\right) \mbox{TR}$.
Having a basis we naturally want to calculate all norms (square). 
ColorMath has special functions for this as well. 
If we define 

\begin{eqnarray}
  \min{OurBasis}= \{ \min{Vector181},\min{Vector818},\min{Vector888s},\min{Vector888a} \}
\end{eqnarray}
we may calculate the squares of the basis vectors using

\begin{eqnarray}
\mdef{CDot[} \mdef{OurBasis} \mdef{]}.
\end{eqnarray}
This results in a \mcom{List} containing the scalar products between each 
basis vector and itself
\begin{eqnarray}
  \left\{
  \mbox{Nc} \left(-1+\mbox{Nc}^2\right) \mbox{TR},\,
  \mbox{Nc} \left(-1+\mbox{Nc}^2\right) \mbox{TR},\,
  \frac{2 \left(4-5\mbox{Nc}^2+\mbox{Nc}^4\right) \mbox{TR}^3}{\mbox{Nc}},\, 
  2\mbox{Nc} \left(-1+\mbox{Nc}^2\right) \mbox{TR}^3\right\}.
\end{eqnarray}

Often non-orthogonal bases are used, in particular, this is the 
case for the trace type bases. In this case, to square a general amplitude,
all the scalar products between all the basis vectors are needed.
For calculating the scalar product matrix the command 
\mcom{CDotMatrix} may be used. This returns 
a \mcom{List} of \mcom{List}s, i.e. a matrix,  where the $ij$-th element 
is the scalar product between the $i$-th and $j$-th vector in
the list of (basis) vectors.
For larger vector spaces with complicated scalar products,
it may be desirable to get progress information on
the calculations. This can be obtained by 
setting the option \mcom{Verbose} $\to$ \mcom{True} for \mcom{CDotMatrix},
\begin{equation}
\mdef{CDotMatrix}[\mdef{OurBasis}, \mdef{Verbose} \to \mdef{True}].
\end{equation}
Similarly, it may be useful to be able to simplify potential (normalization)
roots assuming a large $\mdef{Nc}$.
This can be done by using the option \mcom{NcMin}.
The scalar product functions, along with their options, are listed in 
\tabref{tab:tensor_functions}.

Often it is also of interest to investigate the effect of gluon exchange on
the color structure expressed in a basis, i.e. starting in a basis vector
$j$, what is the effect on the basis vector of exchanging a gluon between
parton1 and parton2. 
This is useful both for soft gluon resummation and for one-loop corrections via
gluon exchange. The result can be expressed in terms of a matrix
whose element $ij$ is the $i$-th component resulting after such
an exchange in the initial vector $j$.
This is calculated by the function
\begin{equation}
\mdef{CGamma}[\mdef{Basis}, \mbox{parton1}, \mbox{parton2}]
\end{equation}
where \mcom{Basis} is a \mcom{List} of basis vectors, defined using the syntax
in \eqref{eq:2qqbarg} and parton1 and parton2 are the numbers of 
the partons in the basis vectors, i.e., in this case numbers between 1 and 5.
The sign conventions are such that quark-gluon vertex always comes without
additional signs, and the triple-gluon vertex have the indices appearing in
the order: external index, internal dummy index and index of the exchanged
gluon\footnote{The sign conventions thus differ from the typical eikonal
choice.}.
For example, we may calculate the effect of
gluon exchange between parton 1 and 3 in the above basis using 

\begin{equation}
\mdef{CGamma}[\mdef{OurBasis},\, \mdef{1},\, \mdef{3}, \,\mdef{Verbose}
\rightarrow \mdef{False},\, \mdef{BasisType} \to \mdef{OrthogonalBasis}].
\end{equation}
Here we have supplied optional information about the basis type,
that the basis is orthogonal, to speed up the calculations.
By default \mdef{CGamma} does a few consistency checks.
It checks that the vector resulting after gluon exchange, when squared,
has the same value as the basis decomposed version.
For orthonormal bases (\mcom{BasisType} $\to$ \mcom{OrthonormalBasis}) it is
also checked that the resulting matrix is symmetric \cite{Seymour:2008xr}.
These checks may, however be turned off  (\mcom{MakeChecks} $\to$
\mcom{False}). The set of options, with default values are
listed in \tabref{tab:tensor_functions}.

\section{Validation and scalability}
\label{sec:validation}

The computational rules and functions in this package have been used for
calculating the three gluon projection operators presented in 
\cite{Keppeler:2012ih}.
This imposes highly nontrivial consistency checks, as it is verified that 
every projector square equals itself, which at intermediate steps
often involve several ten thousand terms.
Additional consistency checks on the color contraction rules have been made by 
using the \mcom{CGamma} function which checks that vectors square and basis
decomposed vectors square agree, and by changing the order in which the color 
structure contraction rules are applied.
The scalar product matrices have also been compared to the ColorFull 
code \cite{Sjodahl:ColorFull} for tree level
trace bases with up to six partons out of which one parton is a quark and one an
anti-quark. Selected results have been compared against
\cite{Sjodahl:2008fz}, and the package has been tested in Mathematica 7,
8 and 9.

The computational effort needed for exact treatment of the color space
grows very quickly with the number of partons. The dimension of the 
vector space grows roughly as a factorial in the number of gluons plus
$\qqbar$-pairs \cite{Keppeler:2012ih} 
(strictly speaking an exponential for finite $\Nc$ in a multiplet basis). The
computational effort for ColorMath, or any program operating by direct
manipulation of quark-lines, tend to grow roughly as the square of this, as the
quark-lines are non-orthogonal.
ColorMath (in its current form) is thus rather intended to 
be an easy to use package for calculations of low and intermediate complexity
than a competitive tool for processes with very many colored partons.

\section{Conclusion}
\label{sec:conclusions}

In this paper a Mathematica package ColorMath for performing color 
summed calculations in QCD is presented. 
This package allows for simple evaluation of QCD color amplitudes which are
expressed in a format which very much resembles how the color 
structure would have been written on paper, see \tabref{tab:basic_objects}. 
The idea is that the user -- for simple cases -- just should give the
expression, and then run 
\mcom{CSimplify}[$\mdef{Expr}$] rather than \mcom{Simplify}[$\mdef{Expr}$]. 
The package is based on advanced pattern matching rules, and a list of rules is
given in \tabref{tab:rules}, whereas functions acting on color
structures are given in \tabref{tab:functions}.

For calculations of intermediate or high complexity it is often
beneficial to use a basis for performing color space calculations.
ColorMath allows for definition of color tensors of form $\mCa:=\ldots$,
carrying an arbitrary set of quark, anti-quark and gluon indices.
Special functions for calculating scalar products,
and investigating the effect of gluon exchange, are given in 
\tabref{tab:tensor_functions}. ColorMath is, however, {\it not}
intended for high speed calculations involving many colored partons.
For this purpose a separate C++ package is written 
\cite{Sjodahl:ColorFull}.

\section*{Acknowledgments}
Terrance Figy, Johan Gr\"onqvist, Simon Pl\"atzer, Stefan Prestel, Johan Rathsman
and Konrad Tywoniuk are thanked for useful comments on the ColorMath code and/or paper.
This work was supported by a Marie Curie Experienced Researcher fellowship of 
the MCnet Research Training network, contract MRTN-CT-2006-035606, 
by the Helmholtz Alliance "Physics at the Terascale" and by the 
Swedish Research Council, contract number 621-2010-3326.

\newpage 

\appendix

\section{Rules and functions}
\label{sec:index_functions}

In this appendix rules and functions for index and color structure probing are
stated. First the basic rules for manipulations are stated in
\tabref{tab:rules}, then the most important functions, operating directly on 
the basic color objects from \tabref{tab:basic_objects} are listed in
\tabref{tab:functions}, whereas special functions for indices and color 
structure probing are stated in \tabref{tab:index_functions}, and 
the functions dealing with color
vector objects are stated in \tabref{tab:tensor_functions}. 

\begin{table}[h]
\caption{\label{tab:rules} The complete set of rules used by ColorMath to
contract indices.
These rules are applied as always in Mathematica, 
by using ``\mcom{Expr}/.\mcom{TheRules}''. 
To ensure that all indices which can be contracted by the rules 
actually are contracted, apply the rules
repeatedly,``\mcom{Expr}//.\mcom{TheRules}''.
Note that all rules have names ending with \mcom{Rules}.}
\begin{tabular}[t]{ |l |  p{12cm} |}
\hline
Rule & Effect \\ [0.5ex] 
\hline \hline
\mcom{AllSimpleRules} & 
\mcom{AllSimpleRules} is the set of all rules involving
$\ColDelta{g1}{g2}$, $\Coldelta{q1}{q2}$, $\Colo{g1}{...}{gk}$ and
$\COLt{g1}{...}{gk}{q1}{q2}$ which do not increase the number of terms, i.e.,
the union of \mcom{SimpleRules} and \mcom{OTSimpleRules}. \\ \hline
\mcom{ExpandThenRules} & 
The rules \mcom{ExpandThenRules} first uses \mcom{Expand},
then applies \mcom{OTThenAllSimpleRules} and finally 
restores default index order with the \mcom{SortIndices} function.
\\ \hline
\mcom{FDRules} & Rules for gluon contraction for terms involving up to three structure constants,
$\Colf{g1}{g2}{g3}$ or $\Cold{g1}{g2}{g3}$, i.e., expressions of form
\eqref{eq:fd1} and \eqref{eq:fd2}.
The terms resulting after contraction are expressed in terms of structure 
constants and gluon deltas, $\ColDelta{g1}{g2}$.
\\ \hline
\mcom{FDToORules} & 
Rules for replacing structure constants, 
$\Colf{g1}{g2}{g3}$ and $\Cold{g1}{g2}{g3}$, with
sums of closed quark-lines $\Colo{g1}{g2}{g3}$ using the intermediate
expression in \eqref{eq:fdMat}.
\\ \hline
\mcom{FDToTRules} & 
Replaces structure constants, $\Colf{g1}{g2}{g3}$ and $\Cold{g1}{g2}{g3}$, 
with a sum of products of SU(\Nc) generators $\Colt{g1}{q1}{q2}$ using the last
form in \eqref{eq:fdMat}. \\ \hline
\mcom{OTGluonRules} &  
Rules for contracting repeated gluon indices in $\Colo{g1}{...}{gk}$ and
$\COLt{g1}{...}{gk}{q1}{q2}$.
The Fierz identity, \eqref{eq:Fierz}, is included in these rules.
\\ \hline
\mcom{OTSimpleRules} & 
Rules for contracting neighboring and next to neighboring gluons 
in closed and open quark-lines, $\Colo{g1}{...}{gk}$ and
$\COLt{g1}{...}{gk}{q1}{q2}$, \eqref{eq:neighbors} and
\eqref{eq:next_neighbors}.
\\ \hline
\mcom{OTThenAllSimpleRules} & 
The rules \mcom{OTThenAllSimpleRules} first applies 
\mcom{OTGluonRules} and then repeatedly \mcom{AllSimpleRules}.
\\ \hline
\mcom{OTToTRules} & Rules for replacing open and closed quark-lines,
$\Colo{g1}{...}{gk}$ and $\COLt{g1}{...}{gk}{q1}{q2}$, with products of
SU($\Nc$) generators $\Colt{g1}{q1}{q2}$. \\ \hline
\mcom{Remove0ORules} & Rule for simplifying closed quark-lines with 
0 gluons, $\colo^{ \{ \} }=\mdef{Nc}$.
\\ \hline
\mcom{Remove0To1ORules} & Rules for simplifying closed quark-lines with 
0 or 1 gluons, $\colo^{ \{ \} }=\mdef{Nc}$, $\colo^{ \{ \blue g1\} }=0$.
\\ \hline
\mcom{Remove0To2ORules} & Rules for simplifying closed quark-lines with 
0 to 2 gluons, $\colo^{ \{ \} }=\mdef{Nc}$, $\colo^{ \{ \blue g1\} }=0$ 
and $\colo^{ \{ {\blue g1},{\blue g2}\} }=\mdef{TR}\, \ColDelta{g1}{g2}$.
\\ \hline
\mcom{SimpleRules}  & 
Basic rules for quark and gluon contraction. These rules involve
$\ColDelta{g1}{g2}$, $\Coldelta{q1}{q2}$ or quark contraction, and never
increase the number of terms. These rules thus contain the rules in (iii-iv) as
well as in \eqref{eq:tr}. \\ \hline
\end{tabular}
\end{table}

\begin{table}[t]
\caption{\label{tab:functions} 
The most useful functions, to be used on any expression carrying color
structure of the form given in 
\tabref{tab:basic_objects}. 
Recall to relabel
dummy indices, manually, or by using \mcom{ReplaceDummyIndices}, and to place
quark
indices that are to be contracted such that one index sits upstairs and one
downstairs. Furthermore, check that the correct \mdef{FullForm} is used, cf.
\tabref{tab:basic_objects}. 
The function \mcom{WhatIsWrong} may be useful for
identifying common input mistakes in the basic color objects.
} \vspace*{0.2 cm}
\begin{tabular}{|p{5.3cm}  |  p{10cm}| } 
\hline
Function with usage & Effect \\ [0.5ex] 
\hline \hline
\hspace*{-0.3 cm}
 \begin{tabular}{l}
    \mcom{CSimplify}[$\mdef{Expr}$]  \\
    \mcom{Options} with default value:\\
    \mcom{RemoveFD} $\to$ \mcom{True}
   \end{tabular}
& 
\vspace*{-7 mm}
The most general function for simplifying color structure. 
If \mdef{\mdef{Expr}} contains structure constants, 
$\Colf{g1}{g2}{g3}$ or  $\Cold{g1}{g2}{g3}$,
\mcom{FDRules} are first applied. 
If, after this, the expression still contains structure constants
they are -- by default -- removed using FDToORules, and all repeated
indices are subsequently contracted
by repeatedly using \mcom{AllSimpleRules}, 
then \mcom{OTThenAllSimpleRules}, and finally \mcom{ExpandThenRules}. 
For color structure containing structure constants, 
it could happen that it is desired not to replace the structure constants.
This can be achieved by setting the option 
\mcom{RemoveFD} $\to$ \mcom{False}.
\\ \hline
\mcom{GluonContract}[$\mdef{Expr}$, $\mdef{Gs}$]  & 
Contracts a set (\mcom{List}) of gluons $\mdef{Gs}=\{\min{g1},...,\min{gn}\}$ 
or a single gluon $\mdef{Gs}=\min{g1}$, in the expression $\mdef{Expr}$, 
while leaving other indices uncontracted. 
This function is intended for quark-lines and will replace 
structure constants with quark-lines.
\\ \hline
\mcom{ReplaceDummyIndices}[$\mdef{Expr}$] & 
Replaces the dummy indices in $\mdef{Expr}$ with 
a new set of unique dummy indices.
\\ \hline
\mcom{SortIndices[$\mdef{Expr}]$} & Sorts the gluon indices appearing in 
$\ColDelta{g1}{g2}$, $\Colf{g1}{g2}{g3}$ or
$\Cold{g1}{g2}{g3}$ and $\COlo{g1}{...}{gk}$ such
that they stand in Mathematica default order.
This is needed to ensure that one color structure only is represented in one
form.
\\ \hline
\mcom{SplitConstAndColor}[$\mdef{Expr}$] & 
Splits an expression $\mdef{Expr}$ into a \mcom{List} of \mcom{List}s 
of color structures and corresponding multiplicative factors 
\{\{constants 1, color structure 1\},\{constants 2, color structure 2\},...\}. 
This is done by first expanding the expression and then splitting the terms
separately.
\\ \hline
\mcom{WhatIsWrong}[$\mdef{Expr}$]& 
Checks if anything is obviously wrong with an expression, 
for example if \mcom{Power} is used instead of \mcom{Superscript} 
or if gluon indices are placed downstairs. 
The check is performed by first expanding the expression, 
and then checking each term. 
Read error messages from above.
\\ \hline
\end{tabular}
\end{table}

\begin{table}[t]
\caption{\label{tab:index_functions} 
The below set of functions are used for probing the color structure, and
identifying indices. Concerning the convention of what (lower or upper)
quark-indices are referred to as incoming quarks and outgoing anti-quarks, as opposed to incoming
anti-quarks and outgoing quarks, ColorMath does not specify any preference,
but simply refers to the indices as upper quarks and lower quarks
when needed.
}
\vspace*{0.2 cm}
\begin{tabular}{ | l  | p{11 cm} | } 
\hline
Function & Effect \\ [0.5ex] 
\hline \hline
\mcom{AllIndices}[$\mdef{Expr}$] & Returns a \mcom{List} of all (external and dummy) indices in \mdef{Expr}.
\\ \hline
\mcom{ContainsColor}[$\mdef{Expr}$] & Returns \mcom{True} if the expression $\mdef{Expr}$ contains any color structure, i.e., any of the
terms in \tabref{tab:basic_objects}, and \mcom{False} otherwise.
\\ \hline
\mcom{ContainsFD}[$\mdef{Expr}$] & Returns \mcom{True} if the expression
$\mdef{Expr}$ contains structure constants and \mcom{False} otherwise.
\\ \hline
\mcom{ContainsGluonDelta}[$\mdef{Expr}$] & Returns \mcom{True} if the expression
$\mdef{Expr}$ contains a gluon delta-function, $\ColDelta{g1}{g2}$, and
\mcom{False} otherwise.
\\ \hline
\mcom{ContainsO}[$\mdef{Expr}$] & Returns \mcom{True} if the expression $\mdef{Expr}$ contains closed quark-lines, 
$\Colo{g1}{...}{gk}$, and \mcom{False} otherwise.
\\ \hline
\mcom{ContainsQuarkDelta}[$\mdef{Expr}$] & Returns \mcom{True} if the expression
$\mdef{Expr}$ contains a quark delta-function, $\Coldelta{q1}{q2}$, and \mcom{False} otherwise.
\\ \hline
\mcom{ContainsT}[$\mdef{Expr}$] & Returns \mcom{True} if the expression $\mdef{Expr}$ contains open quark-lines,
$\COLt{g1}{...}{gk}{q1}{q2}$, and \mcom{False} otherwise.
\\ \hline
\mcom{DummyIndices}[$\mdef{Expr}$] & Finds the dummy indices in an expression $\mdef{Expr}$, 
by first expanding it and then finding all dummy indices in all terms.
\\ \hline
\mcom{GluonIndices}[$\mdef{Expr}$] & Returns a \mcom{List} of all gluon indices (external and dummy) in the expression $\mdef{Expr}$.
\\ \hline
\mcom{LowerQuarkIndices}[\mdef{Expr}] & Returns a \mcom{List} of all (external
and dummy) quark-type indices placed downstairs, i.e., $q2$ in
$\COLt{g1}{...}{gk}{q1}{q2}$ and $\Coldelta{q1}{q2}$.
\\ \hline
\mcom{UpperQuarkIndices}[$\mdef{Expr}$] & Returns a \mcom{List} of all (external
and dummy) quark-type indices placed upstairs, i.e., $q1$ in
$\COLt{g1}{...}{gk}{q1}{q2}$ and $\Coldelta{q1}{q2}$.
\\ \hline
\end{tabular}
\end{table}

\begin{table}[t]
\caption{\label{tab:tensor_functions} Special functions for color
structures expressed in the form of \eqref{eq:2qqbarg}, i.e., the corresponding
tensors need to be defined using the pattern matching underscores, potential dummy indices
should be hidden inside \mcom{Module}s, and the evaluation should be 
delayed for expressions containing dummy indices, i.e. ``:='' 
should be used. 
The vector indices should sit inside a \mcom{List} in the \mcom{Subscript}.
Additional options may be supplied to these functions to control
for example the verbosity. Thus,  \mdef{CNorm} may for example be called using
\mcom{CDot}[\{\min{C1},\ldots, \min{Cn}\}, \mdef{Verbose} $\to$ \mcom{True}],
to get progress information or simply as 
\mcom{CDot}[\{\min{C1},\ldots,\min{Cn}\}] to use the default options.
}
\vspace*{0.2 cm}
  \begin{tabular}{ |p{6.5cm}  |  p{9.1cm} |} 
    \hline 
    Vector function with usage & Effect \\ [0.5ex] 
    \hline \hline
    \begin{tabular}{l}
    \mcom{CDot}[\min{C1}, \min{C2}, \mdef{Options}] \\
    \mcom{Options} with default value:\\
    \mcom{NcMin} $\to$ \mcom{3}
    \end{tabular}& 
    \vspace*{-0.6 cm}
    Calculates the scalar product between two color tensors \min{C1} and \min{C2}. 
    Before using \mcom{CDot} the color tensors $\mCa$ and $\mCb$ should have been defined. 
    For simplifications (in particular of roots) it is by default assumed that
    $\mdef{Nc}$ is at least \mcom{3}. However, this may be
    manually changed by setting the option \mcom{NcMin},
    for example  \mcom{CDot}[\min{C1}, \min{C2}, \mcom{NcMin} $\to$
    \mdef{100}].
    \\ \hline
     \begin{tabular}{l}
    \mcom{CDot}[\{\min{C1},\ldots, \min{Cn}\}, \mdef{Options}] \\
    \mcom{Options} with default value:\\
    \mcom{NcMin} $\to$ \mcom{3}\\
    \mcom{Verbose} $\to$ \mcom{False}
    \end{tabular}& 
    \vspace*{-0.8 cm}
    Similar to \mcom{CDot}[\min{C1}, \min{C2}] but calculates the scalar product
    of a set of color tensors and themselves and returns the result as 
    a \mcom{List} where the $i$th element is 
    \mcom{CDot}[\min{Ci}, \min{Ci}]. 
    In addition to assuming a minimal value of \mdef{Nc} for simplifications, it
    is also possible to get progress information by setting the option
    \mcom{Verbose} to \mcom{True}.
     \\ \hline
    \begin{tabular}{l}
   	\mcom{CDotMatrix}[\{\min{C1},\ldots, \min{Cn}\},\, \mdef{Options}] \\
        \mcom{Options} with default value:\\
    \mcom{NcMin} $\to$ \mcom{3}\\
    \mcom{Verbose} $\to$ \mcom{False}
    \end{tabular}& 
    \vspace*{-0.8 cm}    
    Calculates the scalar product matrix given a \mcom{List} of color
    vectors, i.e., element $ij$ is \mcom{CDot}[\min{Ci}, \min{Cj}]. 
    If the list contains a basis, this function thus returns 
    the scalar product matrix.
    \\ \hline
    \begin{tabular}{l}
   	\mcom{CNorm}[\min{C1}, \mdef{Options}] \\
        \mcom{Options} with default value:\\
    \mcom{NcMin} $\to$ \mcom{3}\\
    \end{tabular}& 
    \vspace*{-0.6 cm}    
    Calculates the norm of a color tensors using \mcom{CDot}.
    For simplifying the result, it may be useful to set the
    option \mcom{NcMin} to a large value. 
    \\ \hline
    \begin{tabular}{l}
   	\mcom{CNorm}[\{\min{C1},\ldots,\min{Cn}\}, \mdef{Options}] \\
        \mcom{Options} with default value:\\
    \mcom{NcMin} $\to$ \mcom{3}\\
    \mcom{Verbose} $\to$ \mcom{False}
    \end{tabular}& 
    \vspace*{-0.8 cm}    
    Similar to \mcom{CNorm} above, but calculates the norms of a \mcom{List} of 
    color vectors \{\min{C1},\ldots,\min{Cn}\}, and returns a \mcom{List}
    containing the norms. Progress information is available by supplying
    the option \mcom{Verbose}$\to$ \mcom{True}.
    \\ \hline
      \begin{tabular}{l}
   	\mcom{CGamma}[\{\min{C1},\ldots,\min{Cn}\}, \mdef{p1},\,\mdef{p2},\,
   	\mdef{Options}] \\
        \mcom{Options} with default value:\\
    \mcom{NcMin} $\to$ \mcom{3}\\
    \mcom{Verbose} $\to$ \mcom{True}\\
    \mcom{MakeChecks} $\to$ \mcom{True}\\
    \mcom{BasisType} $\to$ \mcom{GeneralBasis}
    \end{tabular}& 
    \vspace*{-1.2 cm}  
    Describes the effect of gluon exchange between partons \mdef{p1}
    and \mdef{p2}, on basis vector \min{Cj} as a column vector
    $j$ in a resulting matrix (technically \mcom{List} of \mcom{List}), i.e.,
    element $ij$ is the resulting color structure's $i$-th component.
    The \mcom{BasisType} option should assume one of the values 
    \mcom{GeneralBasis}, \mcom{OrthogonalBasis}, 
    \mcom{OrthonormalBasis} or \mcom{TraceBasis}. The latter 
    is defined to be any basis where the basis vectors consist of 
    {\it one} product of closed and open quark-lines, not a sum.
    By default it is checked that the color tensor, resulting after
    gluon exchange, is the same when squared directly, as when basis
    decomposed, and then squared. This checks that the basis is
    complete and that the basis decomposition is correct. 
    For orthonormal bases, it is also checked
    that the resulting matrix is symmetric. Using \mcom{MakeChecks $\to$ False}
    these checks are turned off.
    By default, progress information is also written out, which can be changed
    with \mcom{Verbose $\to$ False.} \\ \hline
  \end{tabular}
\end{table}

\clearpage

\section*{References}

\bibliographystyle{JHEP} 
 
\bibliography{Refs}

\providecommand{\href}[2]{#2}\begingroup\raggedright\begin{thebibliography}{10}

\bibitem{Hakkinen:1996bb}
J.~Hakkinen and H.~Kharraziha, {\it {COLOR: A Computer program for QCD color
  factor calculations}},  {\em Comput.Phys.Commun.} {\bf 100} (1997) 311--321,
  [\href{http://xxx.lanl.gov/abs/hep-ph/9603229}{{\tt hep-ph/9603229}}].

\bibitem{Alwall:2011uj}
J.~Alwall, M.~Herquet, F.~Maltoni, O.~Mattelaer, and T.~Stelzer, {\it {MadGraph
  5 : Going Beyond}},  {\em JHEP} {\bf 1106} (2011) 128,
  [\href{http://xxx.lanl.gov/abs/1106.0522}{{\tt arXiv:1106.0522}}].

\bibitem{Kuipers:2012rf}
J.~Kuipers, T.~Ueda, J.~Vermaseren, and J.~Vollinga, {\it {FORM version 4.0}},
  \href{http://xxx.lanl.gov/abs/1203.6543}{{\tt arXiv:1203.6543}}.

\bibitem{Paton:1969je}
J.~E. Paton and H.-M. Chan, {\it Generalized {V}eneziano model with isospin},
  {\em Nucl. Phys. B} {\bf 10} (1969) 516--520.

\bibitem{Dittner:1972hm}
P.~Dittner, {\it Invariant tensors in {SU(3)}. {II}},  {\em Commun. Math.
  Phys.} {\bf 27} (1972) 44--52.

\bibitem{Cvi76}
P.~Cvitanovi{\'c}, {\it Group theory for {F}eynman diagrams in non-{A}belian
  gauge theories},  {\em Phys. Rev. D} {\bf 14} (1976) 1536--1553.

\bibitem{Cvitanovic:1980bu}
P.~Cvitanovi{\'c}, P.~Lauwers, and P.~Scharbach, {\it Gauge invariance
  structure of quantum chromodynamics},  {\em Nucl. Phys. B} {\bf 186} (1981)
  165--186.

\bibitem{Mangano:1987xk}
M.~L. Mangano, S.~J. Parke, and Z.~Xu, {\it Duality and multi-gluon
  scattering},  {\em Nucl. Phys. B} {\bf 298} (1988) 653.

\bibitem{Mangano:1988kk}
M.~L. Mangano, {\it The color structure of gluon emission},  {\em Nucl. Phys.
  B} {\bf 309} (1988) 461.

\bibitem{Nagy:2007ty}
Z.~Nagy and D.~E. Soper, {\it Parton showers with quantum interference},  {\em
  JHEP} {\bf 09} (2007) 114, [\href{http://xxx.lanl.gov/abs/0706.0017}{{\tt
  arXiv:0706.0017}}].

\bibitem{Sjodahl:2009wx}
M.~Sjodahl, {\it Color structure for soft gluon resummation -- a general
  recipe},  {\em JHEP} {\bf 0909} (2009) 087,
  [\href{http://xxx.lanl.gov/abs/0906.1121}{{\tt arXiv:0906.1121}}].

\bibitem{Platzer:2012np}
S.~Platzer and M.~Sjodahl, {\it Subleading {$N_c$} improved parton showers},
  {\em JHEP} {\bf 1207} (2012) 042,
  [\href{http://xxx.lanl.gov/abs/1201.0260}{{\tt arXiv:1201.0260}}].

\bibitem{Sjodahl:2008fz}
M.~Sjodahl, {\it Color evolution of 2 $\to$ 3 processes},  {\em JHEP} {\bf 12}
  (2008) 083, [\href{http://xxx.lanl.gov/abs/0807.0555}{{\tt
  arXiv:0807.0555}}].

\bibitem{Seymour:2008xr}
M.~H. Seymour and M.~Sjodahl, {\it {Symmetry of anomalous dimension matrices
  explained}},  {\em JHEP} {\bf 12} (2008) 066,
  [\href{http://xxx.lanl.gov/abs/0810.5756}{{\tt arXiv:0810.5756}}].

\bibitem{Keppeler:2012ih}
S.~Keppeler and M.~Sjodahl, {\it {Orthogonal multiplet bases in SU(Nc) color
  space}},  {\em JHEP} {\bf 1209} (2012) 124,
  [\href{http://xxx.lanl.gov/abs/1207.0609}{{\tt arXiv:1207.0609}}].

\bibitem{Sjodahl:ColorFull}
M.~Sjodahl, {\it {ColorFull -- A C++ package for color space calculations}},
  {\em (work in preparation)}.

\end{thebibliography}\endgroup

\end{document}